\documentclass[rmp,showpacs,twocolumn]{revtex4}

\usepackage[german,english]{babel}
\usepackage{amssymb}
\usepackage{amsfonts}
\usepackage{amsmath}

\begin{document}

\title{When translocation dynamics becomes anomalous}

\author{Ralf Metzler}
\affiliation{NORDITA, Blegdamsvej 17, DK-2100 Copenhagen {\O}, Denmark}
\email{metz@nordita.dk}
\author{Joseph Klafter}
\affiliation{School of Chemistry, Tel Aviv University, Ramat Aviv,
69978 Tel Aviv, Israel}
\email{klafter@post.tau.ac.il}

\date{\today}

\begin{abstract}
Recent single molecule experiments probing the passage process of a short
single-stranded DNA (ssDNA) through a membrane channel (translocation) allow
to measure the passage time distribution. Building on a recent modelling
approach (D. K. Lubensky and D. R. Nelson, Biophys. J. {\bf 77}, 1824 (1999)),
which has been demonstrated to be valid for chains of up to $\simeq 300$
nucleotides and therefore well applies to the system we have in mind,
we discuss the consequences if
the associated dynamics is not of Markov origin, but if strong memory effects
prevail during the translocation. Motivation is drawn from recent results
indicating that the distribution of translocation times is broader than
predicted by simple Markovian models based on Brownian motion.
Running title: Anomalous translocation dynamics
\end{abstract}


\maketitle

The translocation of biomolecules through membrane pores (channels) is one of
the most vital processes within or across biological cells, serving both
delivery and signalling purposes \cite{alberts}. In (bio)chemistry, forced
translocation is used in selection/purification of larger molecules, and in
medicine, it plays an important role in drug delivery. Whereas the
translocation of short, inflexible molecules is primarily determined by the
properties of the pore (energy-driven transport, sticking events
within the pore, etc.) and the difference of the chemical potential between
the cis and trans sides of the pore, semiflexible and flexible molecules, in
addition, have to cross an entropy barrier while being (partially) confined
within the channel \cite{lubensky,muthukumar,muthukumar1,slonkina,park}. In the presence of a high external bias and for the rather
short chains used in typical experiments, the entropic slowdown as well as the
other interactions between chain and channel wall become negligible, the
passage being dominated by the applied drift \cite{lubensky}.
In what follows, we develop a scenario according to which the translocation
dynamics is governed by slowly decaying memory effects, leading to
a different behaviour in the distribution of passage times which we
believe can be measured experimentally.

Experimentally, the translocation of ssDNA can be observed on a single
molecular level, both voltage-driven \cite{kasia,kasia1,meller1} and in the absence
of an external electric field \cite{meller2}. In such single-molecule
translocation assays, fairly short chains are used, with some 60 bases
corresponding to approximately 12 persistence lengths, or 6 Kuhn lengths
\cite{frank}. The width ($\simeq 50${\AA}) of the
membrane amounts to about one persistence length ($\simeq 40${\AA})
of the ssDNA. A good measure for the translocation process is the distribution
of passage times, i.e., the statistics of time spans the chain needs to cross
from the entry (cis) side to the exit (trans) side of the pore. In
the results, one observes two (or three) different time scales: the shortest
corresponds to chains which retract from the pore back to the cis side, before
completing the passage through the pore; the other (one or two) correspond(s)
to real passage times (if there are two peaks, this can be explained by
different orientations of the chain in respect to the passage direction
\cite{lubensky}).

In a recent experiment \cite{meller2}, it was noted that the first passage
time distribution contains non-negligible contributions over a large time-range
even in the presence of a low driving voltage, a case in which a Markovian model
would predict exponentially fast decay. This may well indicate that additional
mechanisms so far neglected play a role in the translocation dynamics, which
might effect long-tailed first passage time distributions, and therefore imply
a possible modelling by assuming a
non-Markovian behaviour of the system. In this note, we construct a framework
in the limit of strong non-Markovian effects, taking into account {\em
anomalous\/} translocation dynamics through long-tailed memory effects. Given
the accuracy of the newly reported experiments in \cite{meller2}, it might well
 be possible to resolve such effects in log-log analyses of the presently
available, or future data. We collect a number of possible sources for such
anomalous dynamics.

In the presence of a bias field and for chains with $\lesssim 300$
nucleotides, the translocation dynamics in the Markov limit has been
shown to follow the Smoluchowski-type equation \cite{lubensky}
\begin{equation}
\label{smolu}
\frac{\partial P}{\partial t}=\left(-v\frac{\partial}{\partial x}+K
\frac{\partial^2}{\partial x^2}\right)P(x,t),
\end{equation}
where $P(x,t)$ is the probability density function (pdf) to find the chain at
position $x$ at time $t$, and $v$ and $K$ are the associated drift and
diffusion constants, which may be determined from more microscopic models
\cite{lubensky}. The translocation process without retraction from the pore
can thus be described by the first passage time distribution $F(t)$
from the point $x=L$ to $x=0$.
In the presence of the external drift, this leads to the result
[see, for instance, \cite{redner}]
\begin{equation}
\label{fptd}
F_1(t)=\frac{1}{\sqrt{4\pi Kt^3}}\exp\left(-\frac{(L-vt)^2}{4Kt}\right).
\end{equation}
Note the sharp exponential decay for longer passage times. In this picture of
Markov advect\-ion-diffusion, the mean first passage time from the origin to
an absorbing boundary a distance $L$ away is given by $T_1
\equiv\int_0^{\infty}tF_1(t)dt=L/v$, i.e.,
the statistical mean corresponds exactly to a classical linear motion
with the drift velocity $v$. Thus, the influence of the diffusivity in
this average becomes negligible and the motion can be characterised by
the mean $\langle x\rangle=vt$.

\begin{figure}
\unitlength=1cm
\begin{picture}(8,5.1)
\put(-2.1,-0.1){\includegraphics{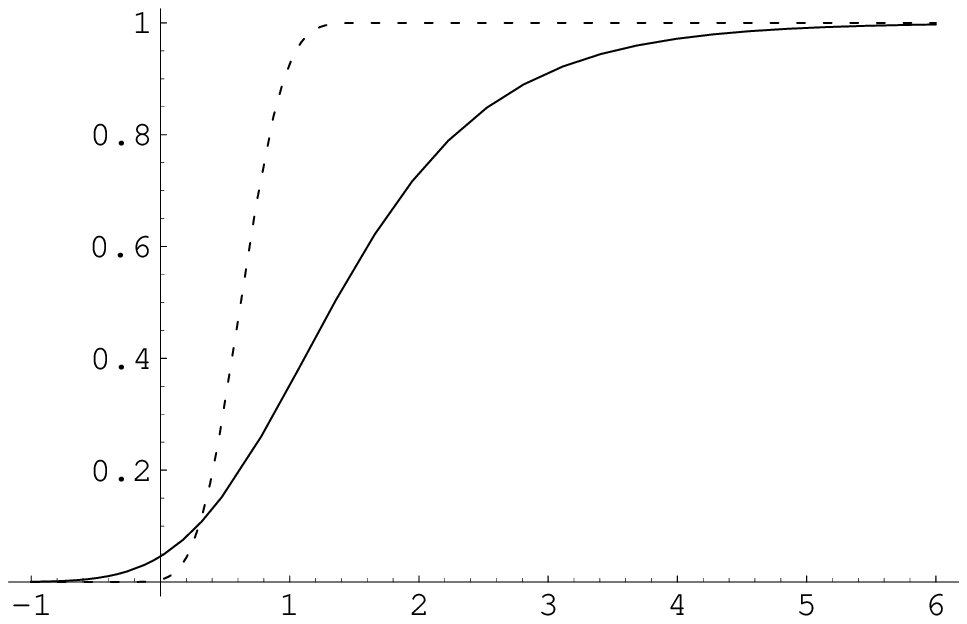}}
\put(0.1,0.5){\includegraphics{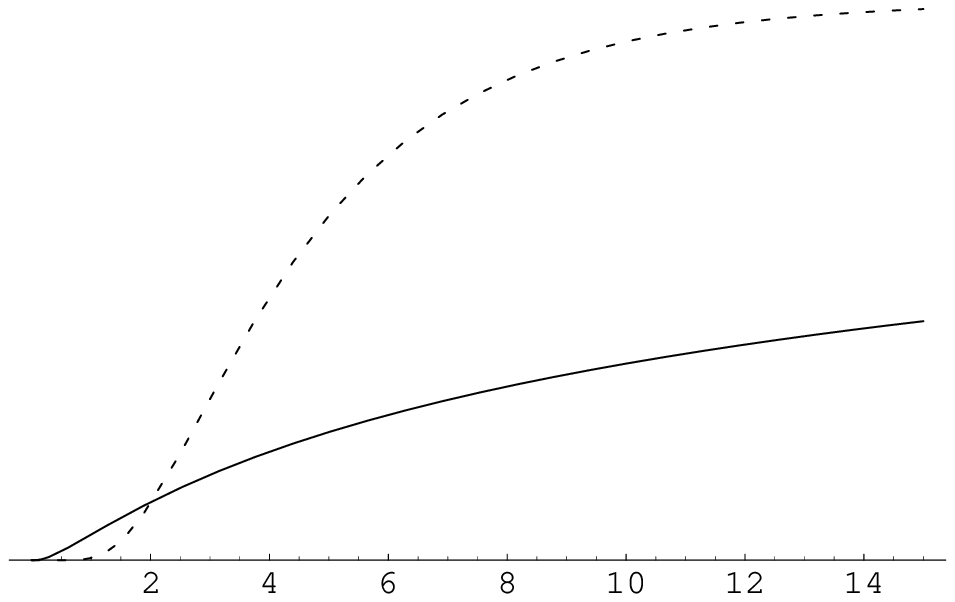}}
\put(7.4,0.6){$t$}
\put(0.8,4.7){$1-S(t)$}
\put(7.3,-0.1){$\log_{10}t$}
\end{picture}
\caption{Translocation probability $1-S(t)$ for long-tailed waiting time wi
th
index $\alpha=1/2$ (full line), in comparison to the Markov counterpart
(dashed) on a log-lin plot. In the inset (lin-lin plot), the much slower
increase of $1-S(t)$
in the non-Markovian result is even more distinct. In the plot, the
dimensionless quantities, $L=5$, and $v=1$ were chosen.
\label{fig}}
\end{figure}

This picture dramatically changes in the presence of long-tailed memory,
effected by a waiting time distribution
\begin{equation}
\label{wtd}
\psi(t)\sim\frac{\tau^{\alpha}}{t^{1+\alpha}}, \qquad (0<\alpha<1),
\end{equation}
according to which interruption times of the transport process in a multiple
trapping model are distributed \cite{report,fkk}, i.e., the transport process
is being stalled successively, and the sticking intervals follow (\ref{wtd}).
In this situation, the relation for the mean gets modified to $\langle x
\rangle=v_{\alpha}t^{\alpha}$, where $v_{\alpha}$ is an appropriately
generalised velocity, and the corresponding generalisation of
equation (\ref{smolu}) is the fractional Fokker-Planck-Smoluchowski equation
\cite{report,fkk}
\begin{equation}
\label{fsmolu}
\frac{\partial P}{\partial t}=\, _0D_t^{1-\alpha}\left(-v_{\alpha}\frac{
\partial}{\partial x}+K_{\alpha}\frac{\partial^2}{\partial x^2}\right)P(x,t),
\end{equation}
with $K_{\alpha}=Kv_{\alpha}/v$, and the fractional Riemann-Liouville operator
$_0D_t^{1-\alpha}P(x,t)=\frac{\partial}{\Gamma(\alpha)\partial t}\int_0^t
\frac{P(x,t')}{(t-t')^{1-\alpha}}dt'$.
The waiting time in expression (\ref{wtd}) is scaled to the
internal time scale $\tau$ \cite{report}.
A typical feauture in this anomalous case is that the maximum of the pdf
due to the strong persistence
of the initial condition (i.e., the probability $\int_0^t\psi(t)dt$ of
{\em not\/} moving), remains at the initial location $x=0$, and the mean
first passage time {\em diverges\/}: $T_{\alpha}=\int_0^{\infty}F_{\alpha}
(t)tdt\to\infty$. The associated
first passage time distribution in the presence of (\ref{wtd}) can be
determined from the classical result (\ref{fptd}). Recalling the scaling
relation for the pdf $P(x,u)$ in Laplace space ($P(x,u)\equiv\int_0^{\infty}
P(x,t)\exp(-ut)dt$) between the solutions of (\ref{smolu}) and (\ref{fsmolu})
\cite{report}, the following scaling holds between the Markov
survival probability $S_1(t)\equiv\int_0^tF(t)dt$ and its non-Markov
analogue,
\begin{equation}
\label{trafo}
S_{\alpha}(u)=u^{\alpha-1}S_1\left(u^{\alpha}\right),
\end{equation}
in rescaled variables. This scaling relation can be rewritten in terms of
the generalised Laplace transformation
\begin{equation}
\label{trafo1}
S_{\alpha}(t)=\int_0^{\infty}E^S(s,t)S_1(s)ds,
\end{equation}
where the Laplace transform of the kernel $E^S(s,u)$
is given by the one-sided L{\'e}vy distribution $E^S(s,u)=u^{
\alpha-1}\exp\left(-su^{\alpha}\right)$, and therefore \cite{report}
\begin{equation}
E^S(s,t)=\frac{1}{s}\sum_{n=0}^{\infty}\frac{(-1)^n}{\Gamma(1-\alpha-\alpha n)
\Gamma(1+n)}\left(\frac{s}{t^{\alpha}}\right)^{1+n}.
\end{equation}
In particular, for $\alpha=1/2$, one obtains $E^S(s,t)=(\pi t)^{-1/2}\exp
\left(-s^2/(4t)\right)$. The pronounced difference between the Markov result
and its non-Markov analogue in the presence of the long-tailed distribution
(\ref{wtd}) for the experimentally measured quantity $1-S(t)$, i.e., the
translocation probability, is displayed in figure \ref{fig}.

\begin{figure}
\unitlength=1cm
\begin{picture}(8,5.4)
\put(0.8,1.6){\includegraphics{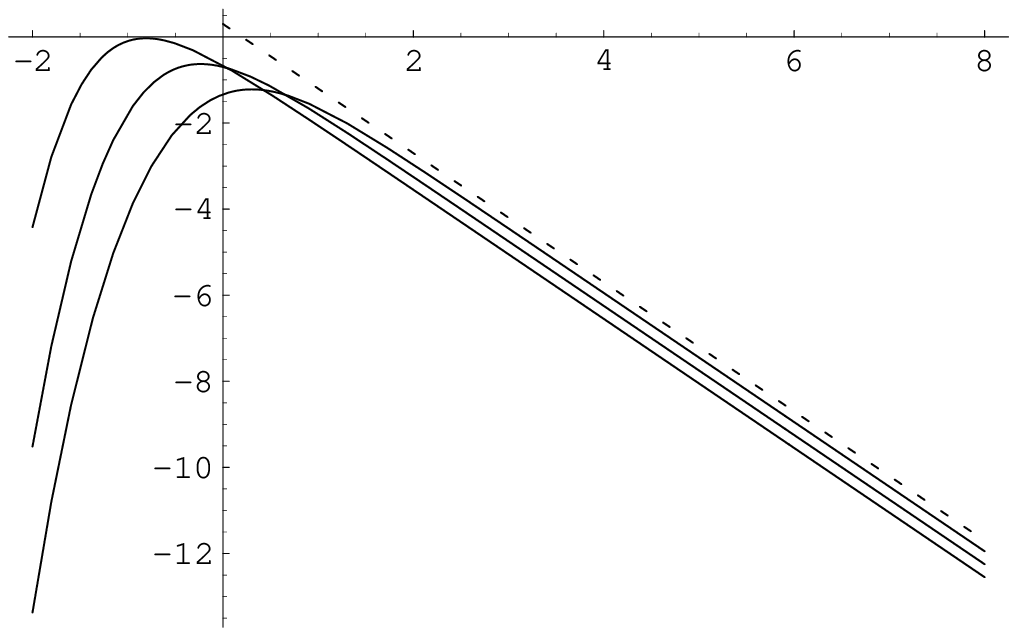}}
\put(-1.8,0){\includegraphics{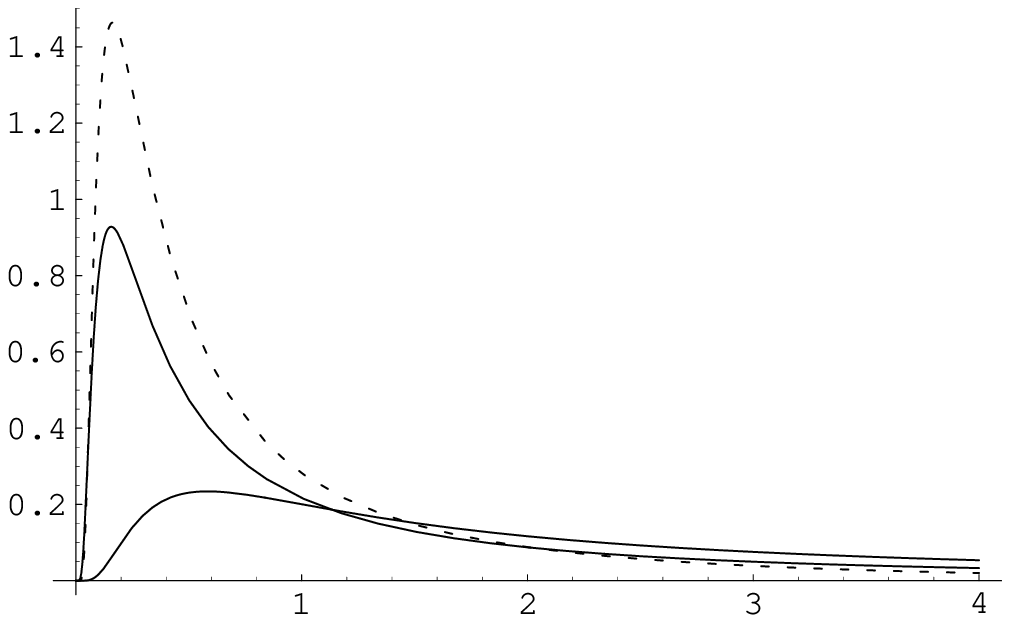}}
\put(8,0.3){$t$}
\put(7.2,4.3){{\small $\log_{10}t$}}
\put(0.35,4.8){$F(t)$}
\put(4,1.8){{\small $\log_{10}F(t)$}}
\end{picture}
\caption{First passage time density $F_{\alpha}(t)$ for $\alpha=1/2$ (full
line; $v=10$, upper curve, and $v=5$) and Markov case (dashed, $v=10$). Not
e
the higher peak in the Markov case, indicating that the
tail falls off faster than in the case with long-tailed waiting times. The
inset shows the double-logarithmic plot with the $-3/2$
power-law asymptotics, for $v=10$, 5, and 2.5.
\label{fig1}}
\end{figure}

It is straightforward to show that
there exists a similar scaling relation for the first passage time density:
\begin{equation}
\label{scal}
F_{\alpha}(u)=F_1(u^{\alpha}).
\end{equation}
From this relation, and the small--$u$ expansion $F_1(u)\sim\exp\left(-Lu/
[2v]\right)$, one can by Tauberian theorems \cite{feller}
deduce the long-time behaviour of the first passage time density,
\begin{equation}
F_{\alpha}(t)\sim t^{-1-\alpha}.
\end{equation}
Thus, in contrast to the Markov case, in the presence of long-tailed
waiting times, the density $F_{\alpha}$ itself exhibits a long-tail, and the
mean first passage time {\em diverges}. In this case, in analogy 
to equations (\ref{trafo}) and (\ref{trafo1}), the transformation
from $F_1$ to $F_{\alpha}$, $F_{\alpha}(u)=\int_0^{\infty}\exp(-u^{\alpha}t)F_1(t)dt$,
can be rewritten as $F_{\alpha}(t)=\int_0^{\infty}E^F(s,t)F_1(s)$, with
\begin{equation}
E^F(s,t)=\frac{1}{t}\sum_0^{\infty}\frac{(-1)^n}{n!\Gamma(-\alpha n)}\left(
\frac{s}{t^{\alpha}}\right)^n.
\end{equation}
For $\alpha=1/2$, one infers $E^F(s,t)=s\left(4\pi t^3\right)^{-1/2}$
$\exp\left(-s^2/(4t)\right)$.\footnote{Note that in the case discussed here no
turnover
between two power-laws, from $t^{\alpha-1}$ to $t^{-\alpha-1}$ results, in
contrast to the properties of the cases discussed in H. Scher and E. W.
Montroll, Phys.  Rev. B {\bf 12}, 2455 (1975); H. Scher, G. Margolin, R.
Metzler, J. Klafter, and B. Berkowitz, Geophys. Res. Lett. {\bf 29},
10.1029/2001GL014123 (2002).}
In figure \ref{fig1}, we depict the functional behaviour of
the first passage time distribution for an external bias, in
comparison to the Markov case, revealing the distinct inverse power-law tails
in the presence of (\ref{wtd}).

Conversely, in the absence of a drift, the reflecting boundary at $x=0$
cannot be a priori neglected (the retraction of the chain
towards the cis side is no longer suppressed by the bias, compare
\cite{lubensky}), and the corresponding first passage time
distribution follows from the eigenvalue problem, such that
the survival probability in the anomalous case is given by a sum over
Mittag-Leffler decays:\footnote{The Mittag-Leffler function $E_{\alpha}(z)
=\sum_0^{\infty}z^n/\Gamma(1+\alpha n)$ is the `natural generalisation' of
the exponential function $e^z\equiv E_1(z)$. $E_{\alpha}(-ct^{\alpha})$
interpolates between an initial stretched exponential and a final inverse
power-law $\sim t^{-\alpha}$ \protect\cite{report}.}
$S(t)=\frac{2}{\pi}\sum_{m=-\infty}^{\infty}\frac{
(-1)^m}{2m+1}E_{\alpha}\left(-\frac{(2m+1)^2\pi^2}{4L^2}t^{\alpha}\right)$,
which in the Markov limit $\alpha=1$ reduce to the exponential function,
and therefore a considerably faster decay
\cite{bvp}. In particular, the long-time behaviour of the first passage
distribution $F(t)\sim t^{-1-\alpha}$ ensues, i.e., the asymptotic behaviour
is analogous to the one derived from the case with drift (in contrast to the
drift-free case without the reflecting barrier, in which $F(t)\sim t^{-\alpha
/2-1}$).

In both cases with and without drift, the first passage time distributions
in the generalised case with long-tailed memory based on the waiting time
form (\ref{wtd}), the corresponding $F(t)$ exhibits a distinct maximum,
its functional form being rather similar to the corresponding Markov 
limit, when viewed on a linear scale. On a double-logarithmic scale,
the power-law $F(t)\sim t^{-1-\alpha}$ differs from the much faster,
exponential decay in the Markov limit. In particular, for both cases
drift-free and with drift, the same power-law behaviour is expected, an
additional consistency check in experiments.
The latter statement, however, only holds if the mechanism effecting
the broad form (\ref{wtd}), and the value of $\alpha$ in particular, is not
affected by the magnitude of the external field. There are indications from
the recent studies \cite{meller2} that this may actually be the case: if the
external drift is increased, the stalling events become less pronounced.
This can influence $\alpha$, but it could also introduce a cutoff at some
time $t_c$ in the waiting time distribution $\psi(t)$, and therefore cause
a system response which is closer to the Markov case than in the absence of
the drift. We stress that in the presence of long-tailed first passage time
distributions, the most probable passage time corresponding to the value at
which $F(t)$ peaks, becomes a rather meaningless quantity, as the mean
first passage time diverges.
We also note that the fit in reference \cite{meller2} of the
passage time distribution by two exponentials with
significantly different time scales in our
approach becomes replaced by a continuum distribution of relaxation times.

Let us now list a number of potential sources for the waiting time
distribution $\psi(t)$:

(i) During the passage, sticking events caused by pore-chain interactions
may occur. As these events would be expected to be correlated with the
co-operative motion of the pore molecules, and possibly the translocating
chain itself, a good guess would be that this causes power-laws of the
form (\ref{wtd}), e.g. as observed in ligand rebinding in proteins \cite{wg}.

(ii) It is known from patch clamp measurements on single fast
chloride channels \cite{blatz}, that ion channels open and close
repeatedly. For the distribution $g(t)$ of duration times of such channel
gating events, it was typically found that it follows a modulated power-law
trend, $g(t)=A(t)t^{-\beta}, \quad \beta>0,$
where the modulation factor $A$ can
follow logarithmic oscillations before eventually
being cutoff by an exponential \cite{nono}, or be {\em constant\/}
\cite{millhauser}. Within a finite time window, both are indistinguishable.
It is therefore fair to say that gating events in a given time window in
single ion channels follow power-law statistics, and typical
values for $\beta$ are around 1.6. The distribution $g$ translates into
our waiting time distribution $\psi(t)$ from equation (\ref{wtd}) with
$\alpha=\beta-1$.

(iii) For longer chains, Chuang et al. argued \cite{chuang}, that the
diffusion of the chain becomes anomalous. Naively viewing the translocation
as a waiting time process during which the monomers in the pore channel
have to wait until they are given way by the vicinal monomers, and so on,
creating a non-Markov process which, on some coarse-grained level, may
well be described by equation (\ref{wtd}), compare also \cite{douglas}.

This list of scenarios is not meant to be complete. However,
one might suspect that the
sticking scenario (i) is most liable to be affected by
the strength of the external bias, producing an effect similar to the
recent experiments reported by \cite{meller2}, in
which the dynamics exhibits the abovementioned turnover from
broad to Brownian motion-type statistics on increase of the external
bias field.

In some translocation experiments, apart from the sharp initial peak in
the first passage time density stemming from immediately retracting chains
back to the cis side there occurs another hump similar to the one of the
translocated chains discussed above. It has been argued that this is due
to the existence of second characteristic passage time, depending on the
orientation of the chain to the membrane channel in respect to the
cis-trans direction (`head or tail first') \cite{lubensky}. The same
effect is expected in the case with long-tailed statistics following
equation (\ref{wtd}). However, it might well be that the associated
power-law exponent $\alpha$ is different for the two orientations,
as the nature of the effective interactions giving rise to the long-tailed
waiting times depend on this head-tail difference.

One might speculate about the biological relevance of anomalous
translocation dynamics. On the one hand, it might be the outcome of
a tradeoff between lack of specificity, if the
passage is too free and a large variety of molecules could pass
the membrane, and too high
suppression, which would require active transport through the
pore, implying a fairly large
energy cost for long molecules. On the other hand, it might be
advantegeous to have a
large variation in the arrival times of translocated molecules
on the trans side (and thereby
very efficient retention of un-translocated molecules on the
cis side).

We have discussed possible changes arising in the distribution of first
passage times in biopolymer translocation through a membrane channel, and
listed a number of reasons which might give rise to such anomalous
behaviour. It should be possible to determine the quantity $F(t)$ from
experiments to sufficient accuracy, in order to be able to distinguish
the normal (Brownian) dynamics result from its anomalous counterpart
in both the presence and absence of an external drift. The large qualitative
difference between exponential and power-law forms should be easily discernible
on a double-logarithmic scale. It should, however, be stressed that the
onset of the power-law trend depends on the strength of the drift, and might
occur for fairly large times if the drift is weak. We finally mention that
the proposed long-tailed effects may also pertain in other systems, like during
the ejection of the DNA of bacteriophages from the capsid through a long pipe-like
channel into the host cell \cite{alberts,muthukumar}.

\acknowledgments

We thank Amit Meller and Ophir Flomenbom for helpful discussions. JK
acknowledges the support of the US-Israel BSF and the Tel Aviv University
Nanotechnology Center.

\end{document}